\documentclass[aps,prb,twocolumn,floatfix,showpacs]{revtex4}
\usepackage{graphicx}
\usepackage{bbm}
\usepackage{amsmath,amssymb}
\newcommand{\be}{\begin{equation}}
\newcommand{\beq}{\begin{equation}}
\newcommand{\ee}{\end{equation}}
\newcommand{\bea}{\begin{eqnarray}}
\newcommand{\eea}{\end{eqnarray}}
\newcommand{\ba}{\begin{array}}
\newcommand{\ea}{\end{array}}

\renewcommand{\t}{\theta}

\begin{document}
\title{Aharonov-Bohm effect in many-electron quantum rings}
\author{V. Kotim{\"a}ki}
\author{E. R{\"a}s{\"a}nen}
\email[Electronic address:\;]{erasanen@jyu.fi}
\affiliation{Nanoscience Center, Department of Physics, University of
  Jyv\"askyl\"a, FI-40014 Jyv\"askyl\"a, Finland}

\date{\today}

\begin{abstract}
The Aharonov-Bohm effect is investigated in 
two-dimensional, single-terminal quantum rings in magnetic fields 
by using time-dependent density-functional theory. 
We find multiple transport loops leading to the oscillation 
periods of $h/(e n)$, where $n$ is the number of loops.
We show that 
the Aharonov-Bohm oscillations are relatively weakly affected 
by the electron-electron interactions, whereas the ring width 
has a strong effect on the characteristics of the 
oscillations. Our results propose that in those experimental 
semiconductor quantum-ring devices that show 
clear Aharonov-Bohm oscillations the electron current is dominated 
by a few states along narrow 
conduction channels.
\end{abstract}


\maketitle

\section{Introduction}

The Aharonov-Bohm (AB) effect is one of the most distinct physical 
phenomena which illustrates the importance of the quantum mechanical 
phase.~\cite{AB1,AB2} In the AB effect, a charged particle acquires 
a phase shift 
\begin{equation}
\phi=2\pi e/h \int_\gamma \mathbf{A}\cdot\mathrm{d}\mathbf{r}
\end{equation}
from the vector potential $\mathbf{A}$ while traveling along the path 
$\gamma$. The existence of this phase shift can be verified, for example, 
by measuring the current through a quantum ring (QR) in a uniform 
perpendicular magnetic field. The electrons traveling along right and 
left arms of the QR gain a relative phase shift
\begin{align} \label{AB_phase_shift}
\phi_r-\phi_l &= \frac{2\pi e}{h}\left(\int_{\gamma_r}\mathbf{A}\cdot\mathrm{d}\mathbf{r}-\int_{\gamma_l}\mathbf{A}\cdot\mathrm{d}\mathbf{r}\right)\nonumber\\
&= \frac{2\pi e}{h}\int_S \mathbf{B}\cdot\mathrm{d}\mathbf{s} = 2\pi\Phi/\Phi_0,
\end{align}
where the surface integral yields the magnetic flux $\Phi$ through the QR, 
and $\Phi_0=h/e$ is the magnetic flux quantum. The resulting current is 
then a periodic function of $\Phi/\Phi_0$. 

The AB oscillation was first observed in 1960 by Chambers 
\emph{et al}.,\cite{Exp0} 
and the first experiment with the period of $\Phi_0$ in a QR 
configuration was made in 1985 by Webb \emph{et al}.\cite{Exp1} 
who measured the 
magnetoresistance of submicron-diameter Au rings. Similar conductance 
oscillations 
were later found also in semiconducting QRs.\cite{Exp2,Exp2b,Exp3,Exp4,Exp4b,Exp8} 
In addition to 
being an important concept in quantum mechanics, the AB effect has emerging 
applications in future computer technologies, random number generation, 
electron 
phase microscopy, and holography.\cite{Applications}

In theory, AB oscillations in QRs have been studied both analytically 
and numerically. 
Analytic works~\cite{Theory1,Theory2,Theory2b,Theory3,Theory4} have focused on 
one-dimensional (1D) rings where the electron path itself is not affected 
by the magnetic field, and thus the situation is different from the 
experiments.\cite{Exp1,Exp2,Exp2b,Exp3,Exp4,Exp4b,Exp8} 
Numerical studies include (single-particle) 
wave-packet simulations in a 1D system,~\cite{Numerics1,Numerics2} and very 
recently also in a two-dimensional (2D) geometry,\cite{Numerics3}
as well as 2D tight-binding calculations for conventional 
QRs~\cite{Numerics4} and graphene rings.~\cite{Numerics5}
These numerical methods are able to take 
the effect of Lorentz force into account, in addition to the AB phase shift 
induced by the vector potential $\mathbf{A}$. The largest effect of the Lorentz 
force is the asymmetric magnetic-field dependent arm injection of electrons which 
leads to the decreasing amplitude of the AB oscillations as the magnetic flux 
through the QR is increased.\cite{Numerics1,Numerics3} The asymmetric arm injection is 
qualitatively taken into an account in a recent S-matrix study by 
Vasilopoulous \emph{et al}.\cite{Theory4} leading to a good agreement with 
numerical results in Refs.~\onlinecite{Numerics1} and \onlinecite{Numerics3}.
In addition to these efforts, the two-particle AB effect has been 
analyzed in rings where the combined path of independent electrons 
enclose the magnetic flux.~\cite{samuelsson}


The aim of this paper is to numerically study the transmission effects in 
a 2D semiconducting QR structure in a static and uniform magnetic field. 
We use time-dependent density-functional theory which allows us to 
examine the real-time dynamics, the electron-electron interactions, and the role 
of the finite ring width (2D character) in the same footing. We find 
that the width of the conduction channel is a critical parameter for the
amplitude of the AB oscillations, whereas the electron-electron interactions 
have a relatively small effect on the oscillation characteristics. The results
give important guidance in assessing the transport dynamics in QR experiments.

The rest of the paper is organized as follows. In Sec. II. we describe the 
QR structure of interest and the theoretical model of the system. The 
computational methods used in this work are presented in Sec. III. The results 
for different number of electrons in the QR 
are presented, analyzed, and compared 
qualitatively with experiments in Sec. IV. The work is summarized in Sec. V.

\section{Model}

We consider a model for a semiconductor QR device fabricated
in AlGaAs/GaAs structures.~\cite{Exp2,Exp2b,Exp3,Exp4,Exp4b,Exp8}
Following the conventional approach in modeling the confined
conduction electrons in the material we apply the effective-mass
approximation with the GaAs parameters, i.e., the effective
mass $m^\ast=0.067\,m_0$ and the dielectric constant 
$\varepsilon=12.7\,\varepsilon_0$. Hence, the energies,
lengths, and times scale as 
$E_{h}^\ast=(m^\ast/m_0)/(\varepsilon/\varepsilon_0)^2E_h\approx 11\,\mathrm{meV}$,
$a_0^\ast=(\varepsilon/\varepsilon_0)/(m^\ast/m_0) a_0\approx 10\,\mathrm{nm}$, and
$t_0^\ast=\hbar/E_h^\ast\approx 60\,\mathrm{fs}$, respectively.
In the results we use these effective atomic units unless stated otherwise.

\begin{figure}
\includegraphics[width=0.80\columnwidth]{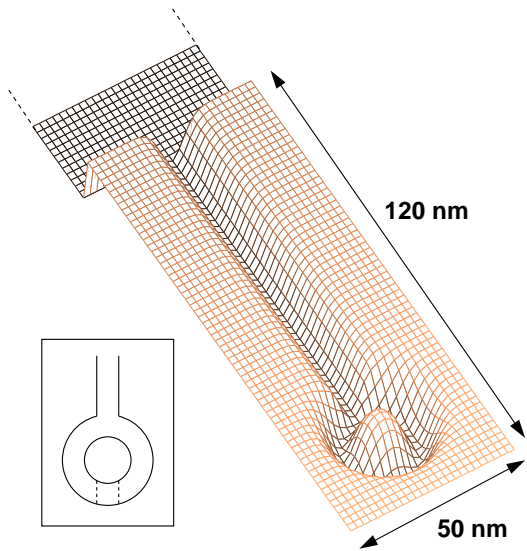}
\caption{(Color online). External potential describing the quantum ring. 
The exact form of the potential can be found in Eq. (\ref{V_ext}). 
Inset: Schematic image of the quantum ring potential, where the 
potential for the initial state ($t=0$) is outlined by the dotted line.}
\label{QR_schematic}
\end{figure}

The time-dependent Hamiltonian describing our $N$-electron QR is given by
\bea \label{Hamiltonian}
\hat{H} & = & {\hat T} + {\hat V}_{ee} + {\hat V}_{\rm ext} = 
\sum_{i=1}^N \frac{1}{2m^\ast}\left[-i\hbar\nabla_i+e\mathbf{A}(\mathbf{r}_i)\right]^2\nonumber \\
& + & \frac{e^2}{4\pi\varepsilon}\sum_{i>j}^N \frac{1}{|\mathbf{r}_i-\mathbf{r}_j|}+\sum_{i=1}^N v_{ext}(\mathbf{r}_i,t),
\eea
where the vector potential is given in the symmetric gauge,
$\mathbf{A}=B/2(-y,x,0)$, for the static and uniform magnetic field 
perpendicular to the QR plane. We neglect the spin effects by
considering spin-compensated systems, so that the Zeeman and spin-orbit 
terms are omitted in the Hamiltonian.
We construct a model for a 
single-terminal QR consisting of $N=1\ldots 10$ electrons and 
represented by the external potential
$v_{ext}(\mathbf{r},t)$. As visualized in Fig.~\ref{QR_schematic},
the potential consists of the QR confinement and the output
terminal. Both the ring and the terminal have a Gaussian-shaped 
cross section and a tunable width (see Appendix).
In addition, we have a linear potential describing a ``bias'' which 
is combined to a rectangular 
potential well at the end of the output terminal. A similar
single-terminal geometry for a metallic QR (in 1D) was considered already in 
Ref.~(\onlinecite{Theory2b}).

The static initial state is calculated for a quantum well
located in the back part of the QR 
(see the inset of Fig.~\ref{QR_schematic}). Then, the
QR and the terminal are suddenly opened up and the electrons
are propagated (at $t>0$) from the quantum well into the QR device.
The opening is described by the Heaviside step function, which
is in fact the only time-dependent part in the external potential 
(and in the Hamiltonian). Details of the potential 
can be found in the Appendix.

Using a single-terminal device guarantees that there are no
undesired effects of electron back-scattering at the input lead. 
To minimize the back-scattering 
at the output of the terminal, the length of the channel 
is relatively large (approximately three times the diameter of the QR).
Moreover, we use a large potential well as a ``sink'' for 
electrons beyond the output of the terminal. The side length of
the rectangular sink is larger than the length of the terminal.

\section{Method}

To examine the time-dependent many-electron system described by the
Hamiltonian in Eq.~(\ref{Hamiltonian}) we apply
time-dependent density-functional theory~\cite{Book-Gross} (TDDFT).
TDDFT is a practical, yet in principle exact approach to
many-body dynamics, and it has received significant
popularity in, e.g., calculating electronic excitations. 
Quantum transport~\cite{diventrabook} can be considered as
a recent field for TDDFT and it has been approached by both 
Green function formalism~\cite{kurth} and finite-system 
time-propagation~\cite{appel_thesis,diventrapaper} -- the latter
being the framework in this study.

TDDFT rests on the Runge-Gross theorem~\cite{Runge-Gross} stating that 
for a given initial state 
there is one-to-one correspondence between time-dependent one-body 
density $n(\mathbf{r},t)$ and the time-dependent one-body potential 
$v_{ext}(\mathbf{r},t)$. It means that a certain time evolution of 
the density is generated by at most one time-dependent potential. Thus, 
it is possible to define an auxiliary Kohn-Sham (KS) system of 
noninteracting electrons moving in time-dependent effective KS 
potential, such that the density of the noninteracting system is {\em equal} 
to the density of the real system. The KS potential consists of
$v_{ext}(\mathbf{r},t)$ (see above),
the Hartree potential corresponding to the classical Coulomb 
interaction, and the exchange-correlation (xc) potential including
all the indirect many-body effects. The xc potential is generally 
non-local in space and time and needs to be approximated 
in practice. Here we apply the common
adiabatic local-density approximation, so that
we {\em locally} and {\em instantaneously} use the
xc potential of the static and uniform 2D electron gas. 
We point out that the ring width is varied within a range 
that does not lead to the recently reported failure of the 
2D local-density approximation in the quasi-1D limit of QRs.~\cite{ring_LDA}

All TDDFT calculations are 
done with the {\tt octopus} code package~\cite{Octopus} published 
under the GPL license. The code has been built on the 
real-space grid discretization method which allows 
realistic modeling of QRs in 2D. The calculations proceed as 
follows: the initial-state configuration is calculated by 
solving the KS equations for the initial external potential 
at $t=0$ and a constant, uniform magnetic field $\mathbf{B}$ perpendicular 
to the plane. The eigenvalues are solved with the 
conjugate-gradient algorithm. After the initial calculation 
the KS orbitals are propagated with the 
external potential at $t>0$ representing the full device 
while keeping the magnetic field constant. The propagation of 
the KS orbitals is done by applying enforced time-reversal 
symmetry approximation for the time-evolution operator 
$\hat{U}(t,t_0)$. After the time-dependent calculation 
the process is started again with a different magnetic 
field strength. We point out that the final propagation
time is limited by undesired back-scattering effects.
Therefore, to obtain a measure for the conductance, we
monitor the electronic density up to a specific 
instant in time as explained below.

\section{Results}

As a measure for the conductivity in our finite system we 
consider the relative amount of transferred electrons
at time $t$ through the ring,
\begin{equation}
N_r(\Phi,t)=1-\frac{N(\Phi,t)}{N},
\end{equation}
where $N=1\ldots 10$ is the total (fixed) number of electrons, 
$\Phi$ is the magnetic flux enclosed by the ring, and $N(\Phi,t)$ 
is the number of electrons in the QR at time $t$, which is 
obtained by integrating the electron density over a 
square which contains the QR. 

\subsection{One-electron transport}

\begin{figure}
\includegraphics[width=0.84\columnwidth]{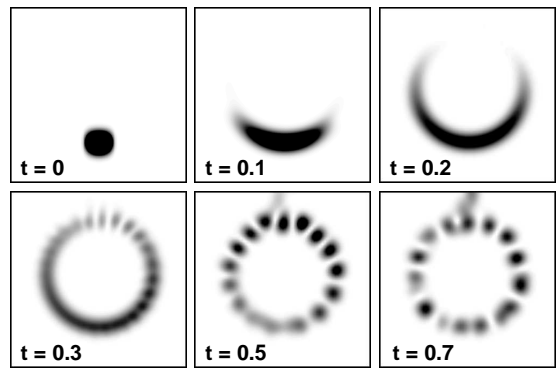}
\caption{(Color online). One-electron density in a QR 
with width parameter $a=0.5$ (see Appendix) at different times.
Magnetic flux through the QR is $\Phi=3\Phi_0$. 
The symmetry of the density distribution is distorted due 
to the Lorentz force.}
\label{one_electron_animation}
\end{figure}

We begin with a single 
electron propagated through the QR. 
As an example of the dynamics of 
the system, the electron density with magnetic flux 
$\Phi=3\Phi_0$ is visualized at 
different times in Fig.~\ref{one_electron_animation}. 
At $t=0$ the density starts to flow through the both 
arms of the QR. The densities in left and right 
arms collide near the output at $t \sim 2$ creating a 
standing wave that oscillates clockwise and counterclockwise
in the system, while a part of the density 
is pumped out from the device. The asymmetric density 
distribution is caused by the Lorentz force.

\begin{figure}
\includegraphics[width=1.00\columnwidth]{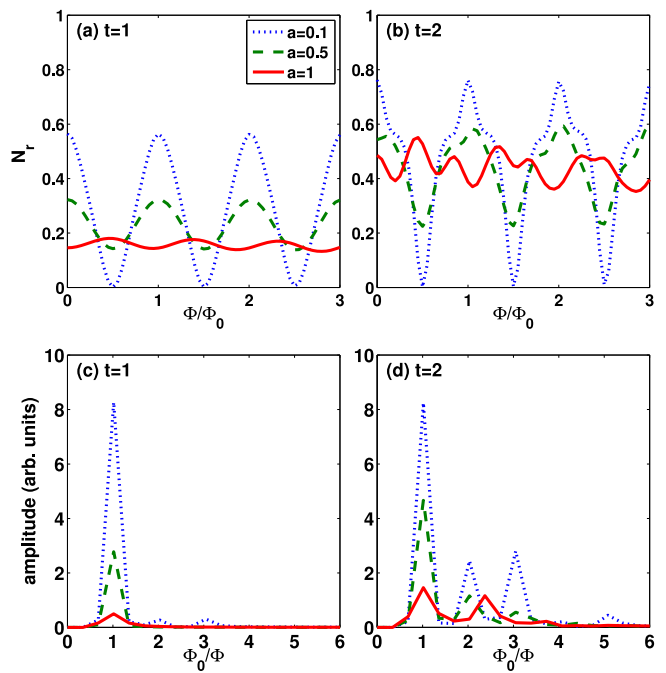}
\caption{(Color online). Transferred probability density for a single-electron 
quantum ring at times (a) $t=1$ 
and (b) $t=2$ as a function of the magnetic flux $\Phi/\Phi_0$. 
(c-d) Corresponding Fourier transforms of the 
situations in (a) and (b), respectively. Larger $a$ corresponds 
to a wider ring.}
\label{One_electron_results}
\end{figure}

The transferred probability density at $t=1$ and $t=2$ is plotted 
in Figs.~\ref{One_electron_results}(a) and (b), respectively.
We have considered three different width parameters for the QR channels,
$a=0.1$, $0.5$, and $1$ (see Appendix),
so that larger $a$ corresponds to larger actual QR width. 
At $t=1$, a smooth and regular AB oscillation
with a period of $\Phi_0$ is clearly visible in the conductance.
The conduction minima and maxima occur at magnetic flux values 
corresponding to phase shifts $\pi$ and $2\pi$, respectively 
[see Eq.~(\ref{AB_phase_shift})]. An exception is the widest
ring with $a=1$, where we find a large, almost a half-period 
phase shift. It is due to transverse density variations 
and large scattering effects in comparison with thinner rings.
In addition, the amplitude of the AB oscillations strongly
decreases as a function of $a$. This is due to the fact that 
the effect of 
destructive interference becomes smaller when there
junction area increases.
In other words, the left and right parts of the electron 
wave packet do not overlap as much as with a thinner output 
terminal. Increasing further the width of the ring and the 
output terminal eventually leads to the vanishing of the 
AB oscillation. Similar destructive effect due to the increasing
QR width has been found by Pichugin~\emph{et al.}~\cite{Numerics4}

As shown in Fig.~\ref{One_electron_results}(b), at $t=2$ 
the regular AB oscillation with the period of $\Phi_0$ is 
distorted by oscillations with larger frequencies.
The Fourier spectra of 
$N_r$ at $t=1$ and $t=2$ in Figs.~\ref{One_electron_results}(c) 
and (d), respectively, show that at first the major 
part of the electron 
transfer occurs at a period of 
$\Phi_0$ in the flux frequency. At larger times the situation changes in an
interesting way, namely, we find additional oscillations
with AB periods of $\Phi_0$, $\Phi_0/2$ and $\Phi_0/3$. 
These oscillations arise from multiple transport loops
in the ring: a part of the electron 
density does not flow out from the QR after traveling 
through one of the arms just once, but it continues to 
flow in the ring and can travel through an odd number 
of QR arms before exiting the ring. A similar phenomenon
has been recently found in tight-binding~\cite{Numerics5} and
wave-packet calculations for QRs.~\cite{Numerics3} 
Multiple transport loops have also been observed experimentally
by Chang {\em et al}.~\cite{Exp8}

Next we analyze the multiple loops in more
detail by considering
the initial current split into two QR arms. The right 
hand part of the current receives a total phase shift
\begin{equation} \label{total_phase_right}
\phi_{R,{\rm total}}(n_r)=\phi_r+\left(n_r-1\right)\left(\phi_r-\phi_l\right),
\end{equation}
where $n_r$ is the number of times the current travels 
through the right arm, and $\phi_r$ ($\phi_l$) is the 
phase shift obtained while traveling through right 
(left) arm once. Similar equation holds for the left 
hand part of the current
\begin{equation} \label{total_phase_left}
\phi_{L,{\rm total}}(n_l)=\phi_l+\left(n_l-1\right)\left(\phi_l-\phi_r\right).
\end{equation}
The total phase difference of the interfering currents 
can be obtained from Eqs. (\ref{AB_phase_shift}), 
(\ref{total_phase_right}) and (\ref{total_phase_left}) as
\begin{equation}
\phi_{R,{\rm total}}-\phi_{L,{\rm total}}=\frac{2\pi\Phi}{\Phi_0/\left(n_r+n_l-1\right)},
\end{equation}
which leads to oscillation with periods of $\Phi_0/n$, where 
$n$ is a positive integer. The amplitude of the flux frequency 
corresponding to the AB period of $\phi_0$ does not 
increase considerably at $t=1...2$: the probability
density traveled through only one arm is small at the 
output of the QR at $t=2$. Proceeding further in time 
leads to more amplitude peaks for flux frequencies 
corresponding to periods $\Phi_0/n$. According to 2D wave-packet 
calculations by Chaves {\em et al.}~\cite{Numerics3}
this effect is not visible in systems having {\em smooth} lead-ring 
connections.

The effect of the Lorentz force is visible 
in the densities plotted in Fig.~\ref{one_electron_animation}. 
The perpendicular magnetic field favors the other arm of the 
QR and produces an uneven distribution of the density in the 
arms. The effect is not that visible in conductance 
graphs in Fig.~\ref{One_electron_results} since our electrons 
start from rest. Szafran \emph{et al.} report {\em increasing 
maximum} of the conductance oscillation together with the 
{\em decreasing amplitude} when the magnetic field is increased:
the Lorentz force bends the path of the incoming electrons so that the 
probability of electron reflection at the input of the 
device decreases. The effect of the Lorentz force is much 
smaller in our simulations, since we do not model the 
scattering effects at the entrance of the QR device.

\subsection{Many-electron transport}

\begin{figure}
\includegraphics[width=1.00\columnwidth]{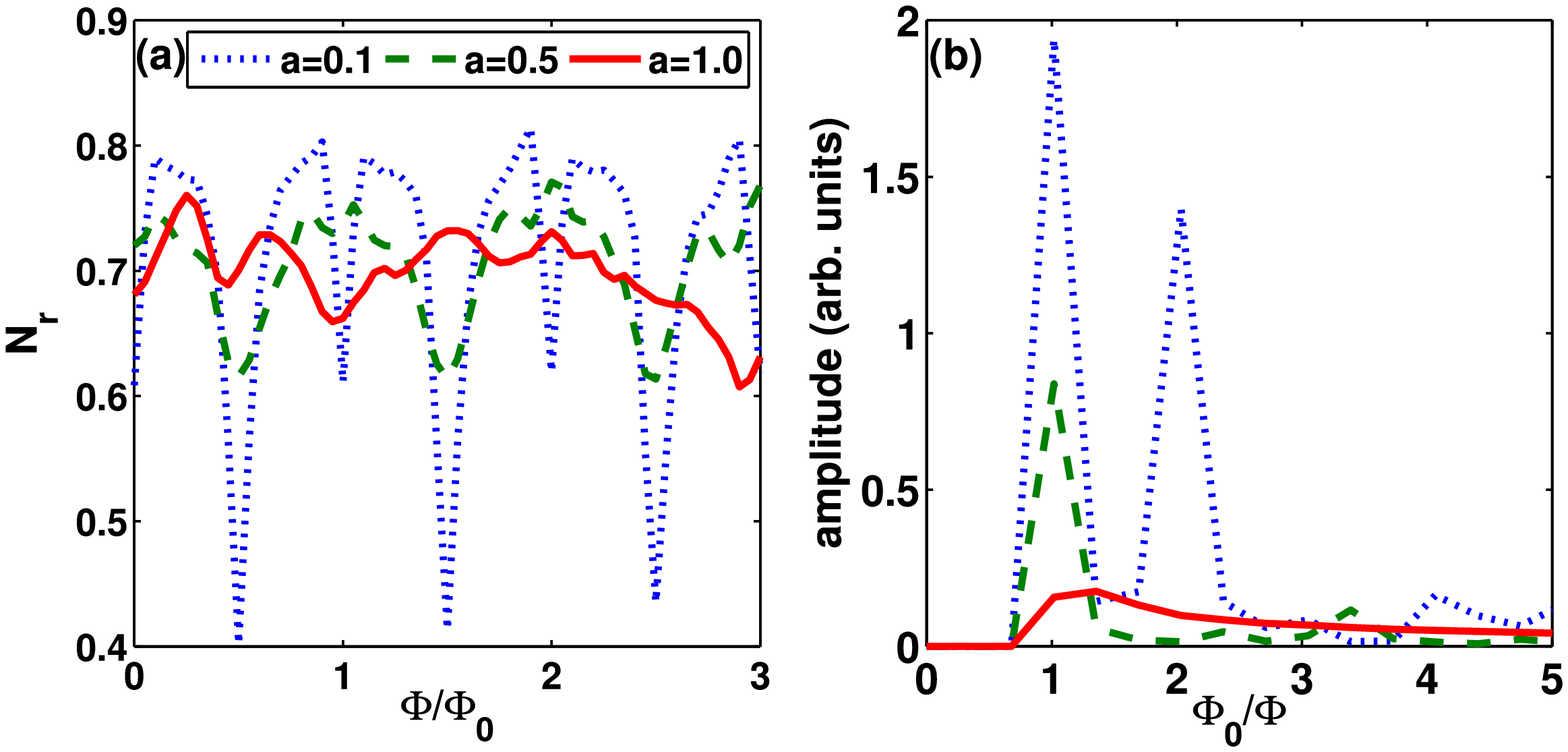}
\caption{(Color online). (a) Transferred relative number of electrons for 
a six-electron quantum ring at time $t=3$ as a function of the magnetic 
flux $\Phi/\Phi_0$. (b) Corresponding Fourier 
transform of the situation in (a). Larger 
$a$ corresponds to a wider ring.}
\label{Six_electron_results}
\end{figure}

Next we consider the electron transport with several 
electrons. The relative transferred number of electrons 
for $N=6$ and the corresponding Fourier spectrum 
are shown in Figs.~\ref{Six_electron_results}(a) and 
\ref{Six_electron_results}(b). 
We find that
increasing the number of electrons does not 
remove the regular AB oscillations, but the system seems
to become more sensitive to changes in the 
width of the ring and the output terminal. 
The sensitivity appears as strong variations in 
$N_r$ as a function of $a$. They are due to complex
electron-electron interaction effects, which, especially
in the case of large $a$, have also significant transverse
contributions. The thinnest ring with width parameter 
$a=0.1$ shows signs of oscillations with phase difference of 
$\pi$ emerging from the interference of scattered and 
unscattered electron densities. In this case the scattering
occurs predominantly along the transport channel.

\begin{figure}
\includegraphics[width=1.0\columnwidth]{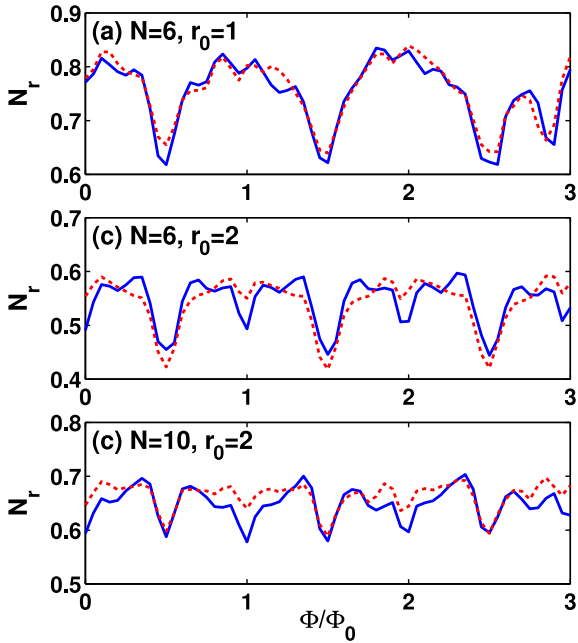}
\caption{(Color online). Transferred relative number of electrons 
for different quantum rings
with (dashes lines) and without (solid lines) interactions.
The width parameter is $a=0.5$ in all cases.
(a) $N=6, \ r_0=1$. (b) $N=6, \ r_0=2$. 
(c) $N=10, \ r_0=2$.}
\label{Many_electron_comparison}
\end{figure}

An explicit view on the effects of electron-electron
interactions is given in Fig.~\ref{Many_electron_comparison},
where we compare the {\em noninteracting} situation (solid lines)
to the interacting one (dashed lines). The conductances are
very similar. This is the case even if the radius of the ring is doubled
[see Fig.~\ref{Many_electron_comparison}(b)], so that
the system is relatively more strongly interacting (note 
that the Coulomb energy scales as $~r^{-1}$, whereas the kinetic
energy scales as $~r^{-2}$). Although the enlargement of the QR 
notably increases the difference between the two situations, 
the qualitative picture remains; most importantly, the AB periodicity remains
the same. However, the interactions seem to make the local minima at 
$\Phi=n\Phi_0$ less pronounced compared to system without 
interactions. This corresponds to a smaller amplitude in 
oscillations with phase shift of $\pi$. 
Generally, the increase in $N$ leads to the damping of the 
oscillation amplitudes as shown in the comparison of the six-
and ten-electron QRs in Figs.~\ref{Many_electron_comparison}(b)
and \ref{Many_electron_comparison}(c). 

Our results above show that the conductance is sensitive to
both the width of the QR as well as the number of electrons
confined in the ring. The effects are, however, different in the
sense that while increasing the width leads to drastic changes
in the AB oscillations even for small $N$, the increase in 
$N$ mainly damps the oscillation amplitudes.
In any case, 
it can be deduced that in semiconductor QRs containing dozens or hundreds
of strongly scattering electrons the AB oscillations are
unlikely to be visible. Hence, in view of the several
experiments on semiconductor QRs where the AB oscillations
are clear,~\cite{Exp1,Exp2,Exp2b,Exp3} we may conclude that in 
a successful transport measurement
the current in the semiconductor 
QR is induced by only the few highest states 
along a narrow, almost ballistic path. In metallic QRs 
(see, e.g., Ref.~\onlinecite{Exp1}) the situation might be
very different.

To qualitatively assess the relative width of the electron path
compared with the QR radius 
we may consider, for example, the experiment of 
Fuhrer {\em et al.},\cite{Exp3} where the ratio between
the width and the radius of the QR device is $\Delta r/r_0 \sim 0.5$. 
To obtain similar AB oscillation amplitudes as in the
experiment for only a {\em single} electron, we need 
to set $\Delta r/r_0 \sim 0.3$ (here to estimate $\Delta r$ we have
cut the density profile at a few percent of its maximum). 
This geometry already corresponds to a significantly thinner ring than
the experimental one, and the difference becomes even larger if
$N$ is increased, since then the width in the calculation
should be decreased to maintain the oscillation amplitudes.
This qualitative analysis suggests that the actual electron
path in the experiment is very narrow compared with the
width of the device itself.

\section{Summary}

We have investigated the Aharonov-Bohm effect in a 
many-electron 
two-dimensional quantum ring with the time-dependent density functional 
theory by modeling the discharging of a one-terminal device. 
We have found multiple transport loops leading
to Aharonov-Bohm oscillation periods of $\Phi_0/n=h/(e n)$, where
$n$ is the number of loops. The Aharonov-Bohm oscillations 
are relatively weakly affected by the electron-electron interactions, 
whereas the ring width has a strong effect on the characteristics of the 
oscillations. In general, the oscillations are notably distorted
when the conduction channels are made wider. Secondly, the
increase in the number of electrons leads to the damping of
the oscillation amplitudes. Our results suggest that 
in experiments on semiconductor quantum rings 
where the Aharonov-Bohm effect is observed,
the actual electron path is narrow compared with the device size, 
and the transport is dominated by a few electrons close to 
the Fermi level.

\appendix

\section{Quantum-ring potential}

The external potential in the Hamiltonian [Eq. (\ref{Hamiltonian})] 
is given by
\begin{equation} \label{V_ext}
v_{ext}(\mathbf{r},t)=\theta(t)V_1(\mathbf{r})+(1-\theta(t))V_0(\mathbf{r}),
\end{equation}
where $\theta$ is the Heaviside step function, $V_0$ is 
the ground-state potential ($t=0$) and $V_1$ is the potential 
used in the time propagation ($t>0$). 
The potentials can be
written as
\begin{align*}
V_{0}(\mathbf{r}) 	&= \Bigl(-V e^{-\bigl(\frac{\sqrt{x^2+y^2}-r_0}{a}\bigr)^2}\\		
		& \times \t(x^2+y^2-(r_0+d/2)^2)\\
		& -\frac{V}{1-e^{-\left(r_0/a\right)^2}} \Bigl(e^{-\bigl(\frac{\sqrt{x^2+y^2}-r_0}{a}\bigr)^2} - e^{-\left(r_0/a\right)^2}\Bigr)\\
		& \times \t((r_0+d/2)^2-x^2-y^2)\Bigr)\\
       		& \times \t(-x) \t (y_{c}^2-y^2)
\end{align*}
and
\begin{align*}
V_{1}(\mathbf{r}) 	&= \min\Bigl(-V e^{(y/a)^2}\t(x),-V e^{-\bigl(\frac{\sqrt{x^2+y^2}-r_0}{a}\bigr)^2}\Bigr)\\
		&  \times \t(L_{\rm pipe}+d/2-x)\t(x^2+y^2-(r_0+d/2)^2)\\
		&  -\frac{V}{1-e^{-\left(r_0/a\right)^2}}\Bigl(e^{-\bigl(\frac{\sqrt{x^2+y^2}-r_0}{a}\bigr)^2}-e^{-\left(r_0/a\right)^2}\Bigr)\\
		&  \times\t((r_0+d/2)^2-x^2-y^2)\\
		&  +V'\bigl(-x/L_{\rm pipe}-V/V'\t(x-L_{\rm pipe}-d/2)\bigr),\\
		& 
\end{align*}
where $V$ is the depth of the potential ring, $V'$ 
is the steepness of the linear potential, $r_0$ is 
the radius of the potential minima of the ring, $y_c$ 
is the cutoff length of initial quantum-well potential 
in the y-direction, $a$ 
is the width parameter of the ring and the channels, $L_{\rm pipe}$ 
is the distance between the center of the ring and the edge 
of the potential well, and $d$ is the grid spacing in the 
simulation box.

Input parameters used in this work are 
$V=300$, $V'=150$, $L_{\rm pipe}=10$, $r_0=1$, $r_0=2$, $a=0.1$, $a=0.5$,$a=1.0$ 
and $d=0.005$. In the propagation we have used time steps
of $\Delta t = 0.0002 \ldots 0.0005$. The length and width of the simulation
box are $45$ and $10$, respectively.

\begin{acknowledgments}
We thank Heiko Appel for the initialization of the present transport
scheme in the {\tt octopus} code and for numerous useful discussions.
This work has been funded by the Academy of Finland. 
In addition, V.K. acknowledges support by the Magnus Ehrnrooth Foundation 
and the Ellen and Artturi Nyyss\"onen Foundation.
\end{acknowledgments}

\end{document}